# Web Search of New Linearized Medical Drug Leads

**Iaakov Exman**

Software Engineering Department
The Jerusalem College of Engineering
POB 3566, Jerusalem, 91035, Israel
`iaakov@jce.ac.il`

Categories and subject Descriptors

H.3.3  [Information Search and Retrieval]: Retrieval models, Search Process {1998 ACM Classification}

Information Systems → Information Retrieval → Retrieval models and ranking
→ Novelty in information retrieval
→ Information Retrieval → Specialized information retrieval
→ Chemical and biochemical retrieval  {CCS 2012}

**Abstract.** The Web is a potentially huge source of medical drug leads. But despite the significant amount of multi-dimensional information about drugs, currently commercial search engines accept only linear keyword strings as inputs. This work uses *linearized fragments* of molecular structures as knowledge representation units to serve as inputs to search engines. It is shown that quite arbitrary fragments are surprisingly free of ambiguity, obtaining relatively small result sets, which are both manageable and rich in novel potential drug leads.

## 1  Introduction

Development of new medical drugs is an extremely long and expensive process (see e.g. [11]). It commonly starts from molecule components with potential desirable activity – called leads (e.g. [3]) – that are gradually improved by adding or modifying groups of atoms that are part of the leads, to increase activity and decrease undesirable side-effects. In pharmaceutical terms, the active groups of atoms that build together the leads are often referred as ligands or just fragments.

Web search is a promising source of leads to new drugs. But, much of the information available about potential drug molecules is either two-dimensional – e.g. planar structural formulas as seen in the next sections – or three-dimensional conformation models. This is an obstacle to search by the generic search engines that are currently limited to one-dimensional keyword strings.

The solution proposed along this work [2] is to characterize drugs by linearized structures that can be sliced into fragments. These fragments can be directly used as search inputs and be stored in suitable ontologies.

The remaining of the paper deals with medical drug knowledge representation issues (section 2), describes specialized medical drug ontologies (section 3), presents preliminary search results (section 4), and concludes with a discussion (section 5).

## 2  Drug Knowledge Representation

Generic search engines are powerful and widely available. They easily deal with textual information.

Medical drugs are based on an active chemical substance, represented by a molecule. Formally, molecules essentially are graphs with edges connecting discrete entities standing for atoms or groups of atoms. Knowledge about a substance contains several information types, of which not all of them can directly serve as input to search engines (see e.g. Konyk at al. [4], and Searls [7]).

Two-Dimensional (2-D) and three-Dimensional (3-D) kinds of information about molecules are amenable to linearization, thus can be useful as search inputs. We mention two of the most common systems of molecular naming that translate structural graph information to compact linear strings.

### 2.1   SMILES

SMILES – an acronym of "Simplified Molecular Input Line Entry" – was proposed by Weininger et al. (see [9], [10]). OpenSMILES [6] is a more recent open standard variant. SMILES unambiguously describes 2-D or 3-D molecular structures using ASCII strings[1]. The latter can be imported by molecular editor tools to convert them back into 2-D or 3-D molecular structures. Thus one manipulates multi-dimensional and linearized structures in practice.

The SMILES string is obtained by a depth-first tree traversal of the molecule graph. Hydrogen atoms implied by the graph skeleton are trimmed. Cycles are broken turning a graph into a spanning tree (numeric suffixes indicate the connected nodes of broken cycles). Parentheses indicate branching points of the tree.

---

[1] Please refer to the literature about SMILES' requirements for unique specification.

Nelarabine – a chemotherapy drug used to treat leukemia – provides a first example of a linearized structure in this paper. Its molecular formula is $C_{11}H_{15}N_5O_5$. Fig. 1 displays a SMILES string for this substance. Fig. 2 shows its 2-D structure.

**COC1=NC(N)=NC2=C1N=CN2C1OC(CO)C(O)C1O**

**Fig. 1**. Nelarabine SMILES String – A linearized string representation of Nelarabine whose two-dimensional structure is seen in Fig. 2. Here one can count e.g. 11 carbon atoms **C** and 5 nitrogen **N** and 5 oxygen **O** atoms. Hydrogen atoms **H** are implicit.

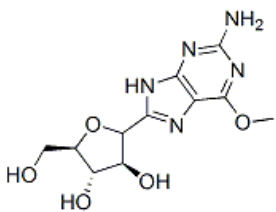

**Fig. 2.** Nelarabine 2-D Structure – This formula displays one hexagonal and two pentagonal rings. Also seen atoms and groups (say $NH_2$=Amino), linked by chemical bonds. In contrast to Fig. 1, some hydrogen atoms **H** are here explicit, while carbon atoms **C** are implied by unnamed vertices of the polygons.

Midazolam – a hypnotic-sedative drug – with molecular formula $C_{18}H_{13}ClFN_3$ is our second example of a linearized structure. Fig. 3 displays its SMILES string. Midazolam's 2-D structure is shown in Fig. 4.

**CC1=NC=C2N1C3=C(C=C(C=C3)Cl)C(=NC2)C4=CC=CC=C4F**

**Fig. 3.** Midazolam SMILES String – Again one can count atom types, except hydrogen H, and compare with its molecular formula. It has one chlorine **Cl** and one fluorine **F** atom.

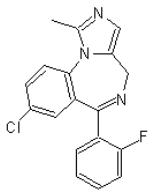

**Fig. 4.** Midazolam 2-D Structure – It shows a pentagonal, two hexagonal and a heptagonal ring. Also explicit some of its atoms (N=Nitrogen, F=Fluor, Cl=Chlorine).

## 2.2   InChi

InChI – an "International Chemical Identifier" – was proposed by the International Union of Pure and Applied Chemistry (IUPAC) in 2006 [5]. It has the same purpose as SMILES.

Each molecular structure has a unique InChI string. The InChI format, algorithms and software are freely available. Search speed is increased by a hashed InChI – the InChIKey – having a fixed 25 character length, a human unreadable form of the InChI descriptor.

InChI describes chemicals in a set of information layers (atoms, connectivity, isotopes, stereo, etc.). A particular InChI descriptor may contain only some of these layers.

SMILES may be less expressive in terms of kinds of information, but it is certainly more human readable than InChI. Therefore we have chosen SMILES for this work.

## 3    Medical Drug Ontologies

Linearized fragments – i.e. whole groups within molecules – are meant to be used as inputs for Lead search. We first concisely review conventional medical drug ontologies, then describe specialized ontologies which store linearized fragments.

### 3.1    Conventional Medical Drug Ontologies

Medical drug and chemical ontologies found in the literature contain detailed and comprehensive descriptions of concepts in the field (e.g. [1], [8]).

The chemical information ontology [1] formulated in OWL (the Web Ontology description Language), precisely defines each of the concepts mentioned in the previous section (molecular formula, structural formula, SMILES, InChi, InChiKey, etc.) among many others.

The Translational Medicine Ontology (TMO) [8] was developed by the W3C (World Wide Web Consortium), within the semantic web for health care. TMO is a patient-centric ontology to integrate data across aspects of drug discovery and clinical practice. TMO is so high-level that it is not really relevant to drug discovery efforts.

### 3.2    Specialized Ontologies for Medical Drug Search

This work proposed specialized "drug-lead" ontologies as repositories of the knowledge acquired about active components in medical drugs.

The drug-lead ontologies are *not* intended to be comprehensive, but rather concise, to facilitate direct and efficient search. Their simple structure has a certain class of drugs – e.g. chemotherapy drugs – as a root. A particular drug instance – say Nelarabine – displays its important components, either by conventional names or linearized structures.

If the important components do not cover the whole molecule, one adds a place-holder termed "skeleton" to notify that the molecular structure contains more components than the explicitly mentioned ones. In other words, skeleton is part of the semantically neutral specification of the structure. This is schematically seen in Fig. 5.

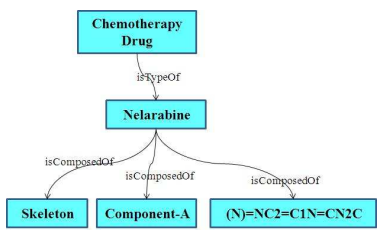

**Fig. 5.** Nelarabine Drug-Lead Ontology – A schematic graph contains the root drug class (Chemoterapy), the Nelarabine instance and its named components – a linearized SMILES substring, a conventional Component-A, and an unspecified skeleton.

Using linearized fragments, say "(N)=NC2=C1N=CN2C" as input to a generic search engine, one actually gets among search results potential leads for the given drug class. – in this case Nelarabine among other chemotherapy drugs – whose structure contains the component.

## 4 Preliminary Results

### 4.1 Experimental Technique

In order to obtain preliminary evaluations for the potential of Web search techniques for drug leads, a set of representative structures of diverse drug classes was chosen. From each representative a random selection of fragments with increasing size was taken and search performed. A few kinds of results were examined, in particular the presence of leads of potential interest (see ref. [2]) and the size trend of the result sets. The latter are shown next.

### 4.2 Search Result Set Sizes

In the next figures one sees search result set sizes for Nelarabine and Midazolam. They are shown as numerical tables and in graphical form.

| Result-set-size | log (result-set-size) | # symbols | fragment |
|---|---|---|---|
| $38.5*10^6$ | 7.59 | 2 | NC |
| $772*10^3$ | 5.89 | 4 | CN2C |
| $189*10^6$ | 8.28 | 6 | =NC2=C |
| $19*10^6$ | 7.28 | 8 | COC1=NC( |
| $9.14*10^3$ | 3.96 | 10 | NC2=C1N=CN |
| 21 | 1.32 | 12 | CN2C1OC(CO)C |
| 3800 | 3.58 | 14 | NC(N)=NC2=C1N= |
| 165 | 2.22 | 16 | (N)=NC2=C1N=CN2C |
| 2540 | 3.4 | 18 | C1=NC(N)=NC2=C1N=C |

**Fig. 6.** Nelarabine Result Sets for Increasing Fragment Size – Fragment size increases from 2 symbols – either atoms or other SMILES symbols (equality or parenthesis) – up to 18 symbols. Search result set sizes are also shown in logarithmic scale.

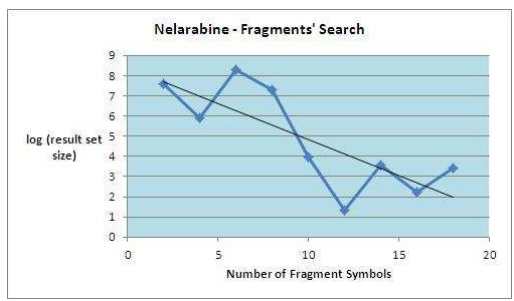

**Fig. 7.** Nelarabine Result Sets for Increasing Fragment Size – Graph showing fragment size increasing from 2 to 18 symbols. Search result set sizes are shown in logarithmic scale. Square (blue) dots are individual results. Continuous (black) straight line is a fitted interpolation.

| Result-set-size | log (result-set-size) | # symbols | fragment |
|---|---|---|---|
| $14.9 \times 10^6$ | 7.17 | 2 | C1 |
| $2.1 \times 10^6$ | 6.32 | 4 | C1=N |
| $229 \times 10^6$ | 8.36 | 6 | N1C3=C |
| $729 \times 10^3$ | 5.86 | 8 | (C=C3)Cl |
| $368 \times 10^3$ | 5.56 | 10 | C(=NC2)C4= |
| $143 \times 10^3$ | 5.16 | 12 | C4=CC=CC=C4F |
| 218 | 2.34 | 14 | C1=NC=C2N1C3)C |
| 3320 | 3.52 | 16 | NC2)C4=CC=CC=C4F |
| 7 | 0.85 | 18 | CC1=NC=C2N1C3=C(C= |

**Fig. 8.** Midazolam Result Sets for Increasing Fragment Size – Fragment size increases from 2 symbols – either atoms or other SMILES symbols (equality or parenthesis) – up to 18 symbols. Search result set sizes also shown in logarithmic scale.

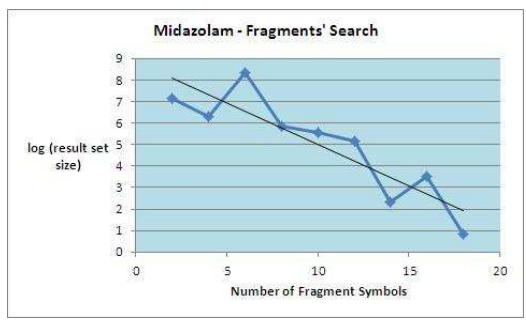

**Fig. 9.** Midazolam Result Sets for Increasing Fragment Size – Graph showing fragment size increasing from 2 to 18 symbols. Search result set sizes are shown in logarithmic scale. Square (blue) dots are individual results. Continuous (black) straight line is a fitted interpolation.

# 5  Discussion

### 5.1 Non-ambiguous Fragments

Three outcomes deserve emphasis among the preliminary results:

a) *Low ambiguity* – since molecular fragments are very uncommon combinations of symbols, not found in natural language, there is virtually no ambiguity of meaning; search results are specific;
b) *Random fragments* – arbitrary slicing of linearized strings, producing random fragments, is surprisingly efficient. There is no need to slice them in the boundary of semantic units. Any slicing will work.
c) *Manageable Result Sets* – the graphs in Fig. 7 and Fig. 9 show the clear trend of reducing result set sizes with increasing fragment size. Thus one can guarantee that above a certain fragment size threshold one gets result sets of the order of hundreds items.

### 5.2   Drug-Lead Ontologies

Knowledge repositories of experience relevant to drug discovery are accepted wisdom. We prefer the specialized compact drug-lead ontologies for Web search, since conventional ontologies are too large and not enough specific, to be efficient.

We have chosen a widely accepted path to new medical drugs – viz. a fragment based approach. The proposed drug-lead ontologies are the vehicle to introduce fragments into search. Linearized components are essential to express structure. The specific choice – to favor SMILES rather than InChi – is not essential and can be changed, if necessary.

### 5.2   Future Work

To demonstrate the efficiency of the taken approach, one needs to make extensive investigation of a variety of drug families.

This work adopted a drug-lead ontology with a dual role of knowledge repository and source of search inputs. A research issue of interest is the number and average sizes of the practical drug-lead ontologies.

### 5.3   Main Contribution

The main contribution of this work is the idea of random fragments of linearized structures for Web search of new medical drugs.

*Acknowledgment*   This work is a continuation of a collaboration with Michal Pinto from the Pharmaceutical Engineering dept. at the Jerusalem College of Engineering.

### References

1. Cheminf = Chemical Information Ontology – in the web site
   http://semanticchemistry.googlecode.com/svn/trunk/ontology/cheminf.owl
2. Exman, I. and Pinto, M.: Lead Discovery in the Web, in Proc. KDIR Conference on Knowledge Discovery and Information Retrieval, Valencia, Spain, (2010).
3. Exman, I. and Smith, D.H.: Get a Lead & Search: A Strategy for Computer-Aided Drug Design', in Symp. Expert Systems Applications in Chemistry, ACS, 196[th] National Meeting, Los Angeles, p. COMP-69, (1988).
4. Konyk, M., A. De Leon, A. and Dumontier, M.: Chemical Knowledge for the Semantic Web, in A. Bairoch, S. Cohen-Boulakia, and C. Froidevaux (Eds.): DILS 2008, LNBI 5109, pp. 169-176, Springer-Verlag, Berlin (2008).
5. McNaught, A.: The IUPAC International Chemical Identifier: InChI  *Chemistry International* (IUPAC) Vol. **28** (6) (2006).
6. OpenSMILES Standard – http://www.opensmiles.org/ Draft (November 2007).
7. Searls, D. B., "Data integration: challenges for drug discovery", Nature Reviews Drug Discovery 4, 45-58 (January 2005).
8. TMO = Translational Medicine Ontology – in web site
   http://translationalmedicineontology.googlecode.com/svn/trunk/ontology/tmo.owl
9. Weininger, D.: SMILES, a chemical language and information system. 1. Introduction to methodology and encoding rules, *J. Chem. Inf. Comput. Sci.* Vol. 28. pp. 31-36 (1988).
10. Weininger, D., Weininger, A., and Weininger, J.L.: SMILES. 2. Algorithm for generation of unique SMILES notation, J. Chem. Inf. Comput. Sci, 29, pp. 97-101 (1989).
11. Wise, M., Cramer, R.D., Smith, D. and Exman, I.: Progress in 3-D Drug Design: the use of Real Time Colour Graphics and Computer Postulation of Bioactive Molecules in DYLOMMS, in J. Dearden, (ed.) Quantitative Approaches to Drug Design, Proc. 4[th] European Symp. on "Chemical Structure-Biological Activity: Quantitative Approaches". Bath (U.K.), pp. 145-146., Elsevier, Amsterdam, 1983.